# Machine-learning-assisted material and geometry characterization from Casimir force measurement

*Hideo Iizuka[1,a) and Shanhui Fan[2,b)]*

[1]Toyota Central R&D Labs., Inc., Nagakute, Aichi 480 1192, Japan

[2]Department of Electrical Engineering, Ginzton Laboratory, Stanford University, Stanford, California 94305, United States

a) hiizuka@mosk.tytlabs.co.jp

b) shanhui@stanford.edu

ABSTRACT: A broadband electromagnetic source is important for scientific and technological applications. Quantum vacuum fluctuations, which manifest most prominently in the Casimir effect, provide a fundamentally broadband electromagnetic source. Here we explore a potential consequence of the broadband nature of quantum vacuum fluctuations, by showing that such fluctuations can enable measurement of material permittivity over a broad frequency range. Specifically, we consider the Casimir force in a parallel-plate geometry, with one plate covered by a nanoscopic thin film. Using a machine learning approach, we show that one can infer both the thickness of the film and its permittivity over a broad frequency range, starting from the dependency of the Casimir forces on the spacing between the two plates. Our work highlights the application potential of using vacuum fluctuations as a naturally-existing broadband electromagnetic source for material characterization, and shows that the inverse problem in Casimir force calculation can be solved with machine learning.

## 1. Introduction

A broadband electromagnetic source is fundamentally important for a wide range of scientific and technological applications, particularly those requiring characterization of optical or material properties over an extended frequency range [1]. Such sources enable measurements of dispersion, absorption, scattering, and nonlinear responses across multiple spectral bands, providing critical information for materials science [2,3], photonics [4-6], and chemistry [7-9]. Commonly used broadband sources include thermal emitters [10,11], such as blackbody thermal emitters, which provide incoherent radiation, as well as supercontinuum laser sources [12,13], which generate spatially and temporally coherent broadband light. Both categories of sources have been extensively deployed in spectroscopy [14,15], imaging [16,17], and metrology [18,19].

In addition to the broadband sources discussed above, quantum vacuum fluctuations provide yet another fundamentally broadband electromagnetic source. These fluctuations correspond to the zero-point motion of the quantized electromagnetic field and possess an extraordinarily wide frequency spectrum, extending continuously from microwave to ultraviolet frequencies and beyond [20]. Quantum vacuum fluctuations manifest in many physical phenomena, most prominently in the Casimir effect [21-27]. In theoretical evaluations of Casimir forces, it is well known that accurate predictions require numerical integration over an exceptionally broad frequency domain in order to achieve agreement with experimental measurements [28-37]. This observation serves as direct evidence of the intrinsically broadband character of quantum vacuum fluctuations.

In this paper, we explore a potential consequence of the broadband nature of quantum vacuum fluctuations, by showing that such fluctuations can function as a broadband source that enables broadband measurement of material permittivity. Specifically, we consider the Casimir force in a parallel-plate geometry, where one of the plates is coated with a nanoscopic thin film. Using a machine-learning methodology [38,39], we show that for such a nanoscopic thin film one can infer both its thickness and its permittivity over a broad range frequency, starting from the dependency of the Casimir forces as a function of the spacing between the two plates. Our work highlights the application potential of using vacuum fluctuation as a naturally-existing broadband electromagnetic source for material characterization, and shows that the inverse problem in Casimir force calculation can be solved with machine learning.

## 2. Configuration and approach

### 2.1. System configuration and Casimir force formalism

We consider a system consisting of two parallel gold plates, where one of the plates is coated with a nanoscopic thin film having permittivity $\epsilon_d(\omega)$ and thickness $t$, separated by the vacuum gap distance $d$, as shown in Fig. 1. Here $\omega$ denotes frequency. The main result of our work is to show that one can obtain the frequency spectra of permittivity $\epsilon_d(\omega)$ and thickness $t$ of the film, from the measurement of the Casimir forces between the plates as gap distance $d$ is varied. We assume non-magnetic isotropic material for the film. The permittivity of the film is given by the N-pole Lorentz-Drude model [28]

$$\epsilon_d(\omega) = 1 + \sum_{n=1}^{N_p} \frac{\omega_{p,n}^2}{\omega_{0,n}^2 - \omega^2 - i\omega\gamma_n}, \tag{1}$$

where $\omega_{p,n}$, $\omega_{0,n}$, and $\gamma_n$ are the plasma frequency, resonance frequency, and damping rate for the $n^{th}$ pole, respectively. The Casimir force in equilibrium $P(d)$ is numerically obtained by the Wick rotation approach [22,26]

$$P(d) = \int_0^\infty k_\| dk_\| \frac{k_B T}{\pi} \sum_{v=p,s} {\sum_{u=0}^\infty}' q_{0,u} \frac{r_{01}(i\xi_u, k_\|)\tilde{r}_{02}(i\xi_u, k_\|)e^{-2q_{0,u}d}}{1 - r_{01}(i\xi_u, k_\|)\tilde{r}_{02}(i\xi_u, k_\|)e^{-2q_{0,u}d}}, \qquad (2)$$

where $k_\|$ and $k_B$ are respectively the lateral wavevector component and Boltzmann constant, and $q_{0,u} = \sqrt{\left(\frac{\xi_u}{c}\right)^2 + k_\|^2}$ with $\xi_u = \frac{2\pi u k_B T}{\hbar}$, $\hbar$ being the reduced Planck constant, and $c$ being the speed of light. $r_{01}(i\xi_u, k_\|)$ and $\tilde{r}_{02}(i\xi_u, k_\|)$ are the Fresnel reflection coefficients from the vacuum (layer 0) to the gold plate (layer 1), and from the vacuum to the film (layer 2) backed by the gold plate (layer 3), respectively, for $v = p$ or $s$ polarization, where $\tilde{r}_{02} = (r_{02} + r_{23}e^{-2q_{2,u}t})/(1 - r_{21}r_{23}e^{-2q_{2,u}t})$ with $r_{ab}$ being the Fresnel reflection coefficient at the interface of layers $a$ and $b$, and $q_{2,u} = \sqrt{\epsilon_d(i\xi_u)\left(\frac{\xi_u}{c}\right)^2 + k_\|^2}$. The prime on the summation denotes that the $u = 0$ term is divided by two. We assume room temperature $T = 300K$. The use of Eq. (2) allows us to obtain the gap dependency of the Casimir force $P(d)$, from the permittivity $\epsilon_d(\omega)$ and the thickness $t$ of the film.

## 2.2. Machine learning approach

We use a machine learning approach to solve the inverse problem of Casimir force calculation, i.e. to obtain the permittivity $\epsilon_d(\omega)$ and the thickness $t$ of the film from the gap dependency of the Casimir force $P(d)$. In our approach, we use a neural network architecture consisting of an input layer ($L_I$-dimensional vector), several hidden layers ($N_L$ layers with each having $N_N$ neurons), and an output layer ($L_O$-dimensional vector). For the input to the neural network, we use $\frac{\partial \tilde{P}}{\partial z}$ at various distances $d$. Here the normalized Casimir force $\tilde{P} = \frac{P}{P_P}$ is the Casimir force $P$ in our system normalized by the Casimir force $P_P = \frac{\hbar c \pi^2}{240 d^4}$ between two plates of perfect electric conductor at the same distance $d$. The use of such spatial derivative of the normalized Casmir force enhances the feature of the distance dependency of the Casimir force. The output of the neural network consists of the thickness of the film, and the pole parameters of the Lorentz-Drude model in Eq. (1).

To train the neural network and to evaluate its performance after training, we first generate a data set consisting of $\frac{\partial \tilde{P}}{\partial z}$ at various distances $d$ for various randomly chosen parameters including the thickness of the film and the pole parameters for the Lorentz-Drude model. We then use part of the data set for training. In each epoch in the training process, we randomly select a batch of incidences from the training data set. The back-propagation algorithm [38] is used to determine weight coefficients between neurons so that the cost function $E = \sum_{m=1}^{M} \sum_{l=1}^{L_o} [log(y_{m,l}) - log(\hat{y}_{m,l})]^2$ is minimized, where $y_{m,l}$ and $\hat{y}_{m,l}$ are the true and predicted $l^{th}$ values in the output layer for the $m^{th}$ incidence in the batch, with $M$ being the batch size. Here in computing the log function the frequencies are normalized to the unit of $rad/s$, and the thickness is normalizd to the unit of $10^3 m$, so that the contributions from the frequencies and the thickness are roughly equal in the cost function. The update of the weights is performed by $N_{TR}/M$ times in one epoch, where $N_{TR}$ is the number of the training data incidences. The training ends when the variation of the weights between epochs is sufficiently small. (Examples of training data in our machine learning architecture are found in Figure S1 and Table S1 in Supplemental Material [40].) After the training is completed, we report the performance of the neural network on the remaining part of the data set, referred to as the testing data set below.

Machine learning has played significant roles in many fields such as speech recognition [41], image recognition [42], and inverse design for electromagnetic [43] and mechanical [44] devices. In the study of Casimir physics, machine learning has been previously used to predict Casimir energies of objects consisting of Dirichlet boundaries of various shapes [45]. The results that we report here, where we use machine learning to solve the inverse problem in Casimir force calculation, have not been previously reported.

## 3. Results and discussion

Below we provide a few examples to illustrate the performance of our approach.

### 3.1. Two-pole Lorentz-Drude model for film permittivities

Firstly, we consider the two-pole Lorentz-Drude model [$N_p = 2$ in Eq. (1)] for the permittivity of the thin film, with a single degree of freedom in the material parameters, i.e. all parameters for the second pole in the Lorentz-Drude model are related to $\omega_{02}$ as $\omega_{p2} = 4\omega_{02}$ and $\gamma_2 = 0.4\omega_{02}$. And the parameters for the first pole are fixed at $\omega_{01} = 0$, $\omega_{p1} = 1 \times 10^{15} rad/s$, and $\gamma_1 = 1 \times 10^{14} rad/s$. We generate a data set of Casimir forces where $\omega_{02}$ is randomly varied in a range of $3 \times 10^{14} rad/s \leq \omega_{02} \leq 1.25 \times 10^{16} rad/s$, and the thickness $t$ is randomly generated in a range of $10nm \leq t \leq 500nm$. Note that the permittivities thus generated satisfy the Thomas-Reiche-Kuhn sum rule, i.e. $\sqrt{\sum_{n=1}^{N_p} \omega_{p,n}^2} \leq 33eV$. Here $33eV$ corresponds to the aggregate dipole oscillator strength in diamond [46]. All other materials have an aggregate dipole oscillator strength that is lower than diamond. (see Figure 3 of the supplemental material of Ref. 47). The permittivity of gold is given by the Drude model having a plasma frequency of $9eV$ and a damping rate of $0.035eV$ [28]. Casimir forces are numerically obtained from Eq. (2) using the parameters described above. The vacuum gap distance is varied in a range of $5nm \leq d \leq 2500nm$ for subsections 3.1 and 3.2 to investigate the underlying physics. In the case of the two-pole model, our data set consists of 1000 incidences. Among these incidences, $N_{TR} = 800$ data incidences are used for training. The hyperparameters used for training is found in Table S2 of Supplemental Material [40]. $N_{TS} = 200$ incidences are used for demonstrating the performance of the trained neural network. We see in Figure 2 that for two different data sets (Figure 2(a) and 2(b) for one dataset and Figure 2(c) and 2(d) for the other), the predicted thickness $t$ and resonance frequency $\omega_{02}$ excellently agree with true values, as evidenced by the fact that all plotted symbols are along a linear line in each panel. The excellent agreement is quantitatively confirmed from root mean square error (RMSE) provided at the right bottom of each panel, which is defined by $\sqrt{\sum_{j=1}^{N_{TS}}[log(y_{j,l}) - log(\hat{y}_{j,l})]^2 / N_{TS}}$ for the $l^{th}$ parameter in the output layer.

The frequency spectra of typical permittivities are obtained from the predicted $\omega_{02}$ using Eq. (1), and compared with true values in Figure 3(a) and 3(b), respectively, which belong to the two different data sets in Figure 2. The predicted permittivities (solid lines) excellently agree with true values (symbols) for the real and imaginary parts of the permittivities. The predicted and the true values of the thickness of the film also agree very well. (The predicted and true resonance frequency $\omega_{02}$ and thickness $t$ for Figure 3(a) and 3(b) are found in Table S3 of Supplemental Material [40].) Therefore, we validate our scheme for predicting the frequency spectra of permittivities and thicknesses of films from the measurement of Casimir forces with variation of distance $d$ for such a simple two-pole model.

### 3.2. Four-pole Lorentz-Drude model for film permittivies

Secondly, we consider the four-pole model of permittivity $\epsilon_d(\omega)$ of films. Unlike the case of the two-pole model above, here the parameters of all the poles are independently and widely varied to illustrate the broadband nature of our approach. The Drude term with $\omega_{01} = 0$ is included, with the contribution widely varied by the plasma frequency $\omega_{p1}$. We generate a data set consisting of 2000 incidences of $t$ and $\epsilon_d(\omega)$ (see the variation ranges of $t$ and pole parameters for $\epsilon_d(\omega)$ in Table S4 of Supplemental Material [40].), and compute the Casimir force behavior for each incidence. We divide this data set into a training data set consisting of $N_{TR} = 1600$ incidences, and a testing data set consisting of $N_{TS} = 400$ incidences. Figure 4(a)-(l) shows the prediction results of the film thickness $t$ and the pole parameters of the four-pole model for the permittivity $\epsilon_d(\omega)$, using the testing data set. (The hyperparameters for the training are provided in Table S5 of Supplemental Material [40].) In general, we see that the prediction of the thickness is always quite accurate. [Figure 4(a)]. The results also show reasonable agreements between the predicted and true resonance frequencies ($\omega_{02}$, $\omega_{03}$, $\omega_{04}$) [Figure 4(d),(g),(j)] and plasma frequencies ($\omega_{p1}$, $\omega_{p2}$, $\omega_{p3}$, $\omega_{p4}$) [Figure 4(b),(e),(h),(k)], with better agreement obtained for higher frequency resonances. On the other hand, the damping rates ($\gamma_1, \gamma_2, \gamma_3, \gamma_4$) are not predicted well [Figure 4(c),(f),(i),(l)]. The results here on the damping rate prediction is consistent with Refs. 48,49, which showed that in general Casimir forces are less sensitive to material loss.

Figure 5(a)-(c) shows three examples of predicted permittivities from the results of Figure 4 using Eq. (1). The predicted permittivities (solid lines) agree with the true ones (symbols) in the high frequency range in the three panels. (The predicted and true poles for $\epsilon_d(\omega)$ and thickness $t$ for Figures 5(a)-5(c) are found in Table S6 of Supplemental Material [40].) We observe good

agreement across more than two orders of magnitude in frequency spanning from infrared to ultraviolet frequency ranges, showing that our approach indeed can be used to obtain broadband permittivity response. In the low frequency range, there are significant discrepancies between the predicted and true permittivities.

We provide a brief discussion of the physics basis that underlies our machine learning approach. In general, the dependency of the Casimir force on $\epsilon_d(\omega)$ is rather complex [50]. (See Figure S2(a) and S2(b) of Supplemental Material [40].) There are certain heuristic arguments that one can make about such a dependency. In the parallel-plate system with a vacuum gap distance $d$, there is the cutoff frequency $\omega_c = \frac{2\pi c}{d}$. The Casimir force is independent of the frequency response of the permittivity above the cutoff frequency [51]. For example, $5nm \leq d \leq 2500nm$ corresponds to $3.77 \times 10^{17} rad/s \geq \omega_c \geq 7.54 \times 10^{14} rad/s$. As we vary the gap distance $d$, the Casmir force is affected by different part of the permittivity spectrum. Thus, there is information about the permittivity spectrum in the behavior of the Casimir force as a function of the gap distance. The distance dependence of the cut-off frequency provides the physics basis that underlies our approach. Nevertheless, such a heuristic argument alone is not sufficient to allow one to determine $\epsilon_d(\omega)$ from the Casimir force measurement. Our results indicate that the use of neural network in fact can extract sufficient information from the Casimir force measurement to determine $\epsilon_d(\omega)$. These results are not a priori obvious and point to the potential of using neural network in the study of Casimir physics.

We also comment on the physics reasoning that explains our difficulty in obtaining low-frequency permittivity. At room temperature, the Casimir force is in general less sensitive to the material permittivity at low frequencies. This can be understood in the Casimir force formula [Eq. (2)], where the integration is in the form of $\int_0^\infty k_\parallel dk_\parallel$. In the absence of material resonances which provide surface states, the integrand is significant only when $k_\parallel < \frac{\omega}{c}$. Consequently the permittity at frequencies substantially below the material resonant frequencies do not contribute significantly to the Casmir force. As an illustration of the observation here, in Ref. 23, the Casimir force between silicon carbide plates has been investigated, where surface resonances occur in the ultraviolet and infrared regions. It was noted that the surface resonance in the ultraviolet region provided the dominant contribution to the Casimir force at a vacuum gap distance of $10nm$. Therefore, it becomes difficult to infer the low-frequency permittivity from the data of the Casimir force.

### 3.3. Film permittivities including silicon

Thirdly, we examine the applicability of our scheme to realistic materials. In general, table data of measured permittivities for most materials can be fitted by the N-pole Lorentz-Drude model, and such fitting has been widely used in electromagnetic computation. Here, we use the four-pole model, as in Subsection 3.2, but generate another data set of film permittivies with $\omega_{01} \neq 0$, where one testing data incidence corresponds to the measured permittivity of silicon in Ref. 52. (The variation ranges of $t$ and pole parameters for $\epsilon_d(\omega)$ are found in Table S7 of Supplemental Material [40].) For gap distances in experimental studies, Casimir forces were measured in a range of $500nm - 3\mu m$ in the parallel plate structure [34] and $30nm - 300nm$ in the sphere-plate structure [37]. In the parallel-plate structure, radiative heat transfer was measured in a range of $25nm - 8\mu m$ [53]. Based on the distance ranges above, here, a gap distance range of $50nm \leq d \leq 2500nm$ is selected in the third case. Accordingly, the film thickness $t$ is randomly generated in a range of $100nm \leq t \leq 500nm$. Similarly, the Casimir force behavior is computed for each of the generated 2000 incidences of $t$ and $\epsilon_d(\omega)$. Then, the data set is devided into a training data set consisting of $N_{TR} = 1600$ incidences, and a testing data set consisting of $N_{TS} = 400$ incidences. Using the same process in Subsection 3.2, the film thickness $t$ and pole parameters for $\epsilon_d(\omega)$ are predicted. (The prediction results and the hyperparameters are found in Figure S3 and Table S8 of Supplemental Material [40].)

Figure 6(a)-(c) shows three examples of predicted permittivities, with Figure 6(a) corresponding to silicon. We see that the predicted permittivities (solid lines) agree with the measured one (crosses) in Ref. 52 in Figure 6(a), and true values in Figure 6(b) and 6(c). (The predicted and true poles for $\epsilon_d(\omega)$ and thickness $t$ for Figure 6 are provided in Table S9 of Supplemental Material [40].) Therefore, we have verified that our prediction scheme can work for a realistic material. Note that discrepancies occur in the lowest frequency pole (around $1 \times 10^{15} rad/s$) in Figure 6(b) and 6(c) as we discussed in the previous subsection.

### 3.4. Denoising autoencoder

Casimir force measurements are affected by instrumental and background noise and calibration errors[54]. In many fields, autoencoders [55] are widely used to minimize noise effect by encoding and decoding functions, i.e. an input data set is mapped to latent space and then reconstructed in the output layer. Here, we adopt a denoising autoencoder [56] of Figure 7 to minimize noise effect in Casimir force values, as the preprocessing for the prediction scheme of Figure 1. The autoencoder architecture consists of an input layer, several hidden layers, and an output layer, with the features of the middle hidden layer having a reduced number of neurons and the architecture being symmetric. In out study, we assume that noise is added in the Casimir force values, i.e. noisy data incidences $\boldsymbol{X_{noisy}}$ include clean (target) Casimir forces $\boldsymbol{X}$ and noise. The neural network is trained so that the Casimir forces $\widehat{\boldsymbol{X}}$ in the output layer can be close to the clean (target) Casimir forces $\boldsymbol{X}$, by employing cost function $E = \sum_{m=1}^{M} \sum_{l=1}^{L_o} (x_{m,l} - \hat{x}_{m,l})^2$, where $x_{m,l}$ and $\hat{x}_{m,l}$ are the clean (target) and predicted $l^{th}$ Casimir force values for the $m^{th}$ incidence in the batch having size $M$.

The same $N_{TR} = 1600$ training data incidences and $N_{TS} = 400$ testing data incidences as those in the third case in Subsection 3.3 are, respectively, used for training the autoencoder of Figure 7 and evaluating the denoising performance. For proof of concept, Gaussian noise with a standard deviation of 0.02 and zero mean is added. Figure 8(a)-8(c) shows the Casimir forces $\frac{\partial \tilde{P}}{\partial z}$ (note $\tilde{P} = \frac{P}{P_P}$ with Casimir force $P$ and $P_P = \frac{\hbar c \pi^2}{240 d^4}$) with variation of gap distance $d$, which correspond to permittivity spectra of Figure 6(a)-6(c). We see that from noisy Casimir forces $X_{noisy}$ (green line), the smooth lines $\widehat{X}$ (blue) are predicted in the three panels, which agree with the clean (target) Casimir forces $X$ (pink lines) excellently in Figure 8(b) and 8(c) and reasonably in Figure 8(a). (The hyperparameters for the denoisisng autoencoder of Figure 7 are found in Table S10 of Supplemental Material [40].)

Figure 8(d)-(f) shows the permittivity spectra predicted from the corresponding denoised Casimir forces $\widehat{X}$ of Figure 8(a)-8(c). The predicted permittivity spectra (solid lines) reasonably agree with the true values (circles) in Figure 8(e) and 8(f). In Figure 8(d), the permittivity spectrum (solid lines) is blue-shifted due to the discrepancy of the denoised Casimir forces (blue line) from the clean ones (pink line) observed in Figure 8(a). (The predicted and true poles for $\epsilon_d(\omega)$ and thickness $t$ from the denoised Casimir forces are found in Table S11 of Supplemental Material [40].) We validate that the denoising autoencoder can reduce noise effect as the preprocessing for the material and geometry characterization from Casimir force measurement although there is room for improvement of the prediction accuracy in the denoising autoencoder.

### 3.5. Discussion

There are several potential approaches that can further improve the performance of our scheme of permittivity measurement. In our calculation we consider only Casimir forces that arise from quantum fluctuations. To improve the determination of low-frequency permittivity, one may consider thermal and non-equilibrium Casmir forces. As the permittivy varies, these forces show significant variation at relatively large gap spacing. For example, with the right choice of materials, Casimir forces can be repulsive beyond $4 \mu m$ [57,58]. Therefore, the low-frequency permittivity plays a significant role in the behavior of these forces. Conversely, one should be able to use such force behavior to infer low-frequency permittivity. To determine the material permittivies and denoise Casimir forces, we have used the basic back-propagation algorithm [38], and the prediction performance may be improved by sophisticated optimization methods such as Adam [59] and L-BFGS [60].

In this paper, we have considered a parallel-plate geometry. Based on the physics argument above, however, our approach should be applicable for other geometries of Casimir force measurement. In Ref. 61, Casimir forces were measured in a system consisting of a gold coated sphere and a flat gold substrate with a molecular film on top as a function of the vacuum gap distance, for evaluating five different self-assembled bio and organic monolayer thin films. Our scheme should be applicable in such sphere-plate geometry to determine the properties of the thin films.

The four-pole model for $\epsilon_d(\omega)$ has been used for considering silicon as a semiconductor material in our study. It would be interesting to explore other realistic materials having more complex frequency spectra by increasing the number of poles.

## 4. Conclusions

We have shown that quantum vacuum fluctuations can function as a broadband source, enabling the determination of material permittivity over a broad frequency range. Numerical results revealed that the permittivity more than two orders of magnitude in frequency spanning from infrared to ultraviolet frequency range, as well as the thickness of the film were determined through the use of machine learning. Our results point to the availability of vacuum fluctuations as a naturally-existing broadband electromagnetic source for material and geometry characterization, and show that the inverse problem in Casimir force calculation can be solved with machine learning.

**Acknowledgements**

H. I. would like to thank Dr. Hirotaka Kaji for helpful discussion on machine learning. Prof. Fan's contribution to this publication was as a consultant, and was not part of his Stanford duties or responsibilities.

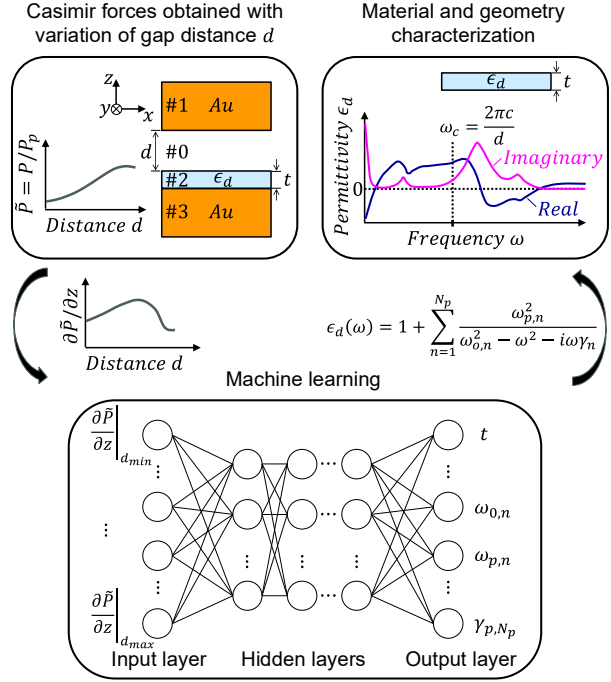

**Figure 1.** Material and geometry characterization from Casimir force measurement. We consider a geometry consisting of two gold regions, with one of the gold regions coated with a dielectric thin film. The two regions are separated by a vacuum gap $d$. The permittivity $\epsilon_d(\omega)$ and the thickness $t$ of the thin film are determined using a neural network. The input to the neural network is the Casimir forces as a function of vacuum gap distance $d$. The neural network outputs the thickness $t$ and the pole parameters of the Lorentz-Drude model for the permittivity $\epsilon_d(\omega)$ [Eq. (1)], from which $\epsilon_d(\omega)$ can be determined.

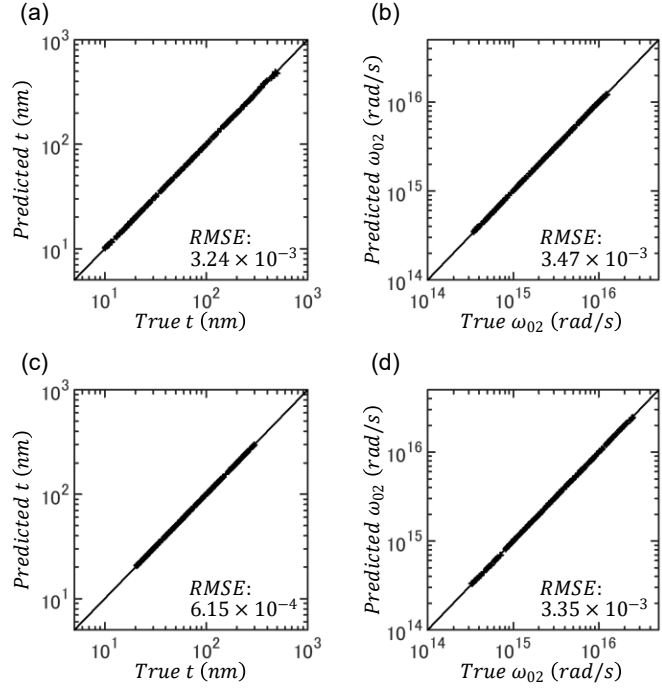

**Figure 2.** Predicted film thickness $t$ and resonance frequency $\omega_{02}$ in the two-pole model for $\epsilon_d(\omega)$ in Eq. (1). 200 testing data points are plotted in each panel. The fact that these data points are along the linear line indicates the agreement between the predicted values and the true values. RMSE is given at the right bottom of each panel. (a) Film thickness $t$ ($10nm \leq t \leq 500nm$) and (b) resonance frequency $\omega_{02}$ ($3 \times 10^{14} rad/s \leq \omega_{02} \leq 1.25 \times 10^{16} rad/s$) for pole parameters of $\omega_{01} = 0$, $\omega_{p1} = 1 \times 10^{15} rad/s$, $\gamma_1 = 1 \times 10^{14} rad/s$, $\omega_{p2} = 4\omega_{02}$, and $\gamma_2 = 0.4\omega_{02}$ in Eq. (1). (c) Film thickness $t$ ($20nm \leq t \leq 300nm$) and (d) resonance frequency $\omega_{02}$ ($3 \times 10^{14} rad/s \leq \omega_{02} \leq 2.5 \times 10^{16} rad/s$) for pole parameters of $\omega_{01} = 0$, $\omega_{p1} = 5 \times 10^{14} rad/s$, $\gamma_1 = 2 \times 10^{14} rad/s$, $\omega_{p2} = 2\omega_{02}$, and $\gamma_2 = 0.8\omega_{02}$.

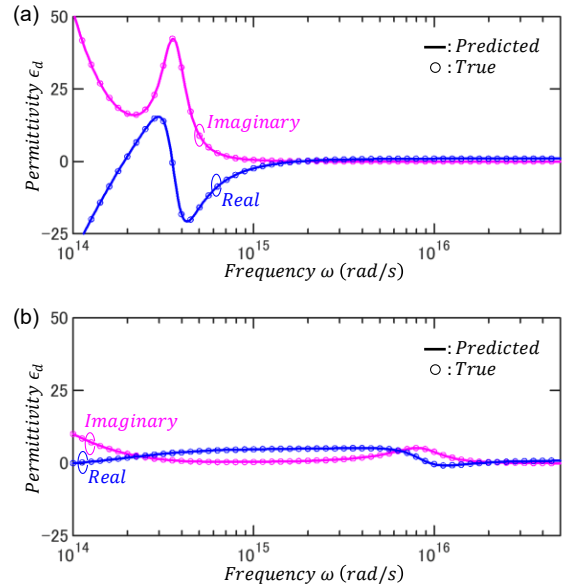

**Figure 3.** (a),(b) Permittivity spectra for the two-pole model. Two examples are selected from each of the two testing data sets, corresponding to Figure 2(a),(b) and Figure 2(c),(d), respectively. The blue and pink solid lines represent the real and imaginary parts of predicted $\epsilon_d(\omega)$, and blue and pink symbols represent the real and imaginary parts of true $\epsilon_d(\omega)$. The pole parameters in Eq. (1) for (a) and (b) are given in Table S3 of Supplemental Material [40].

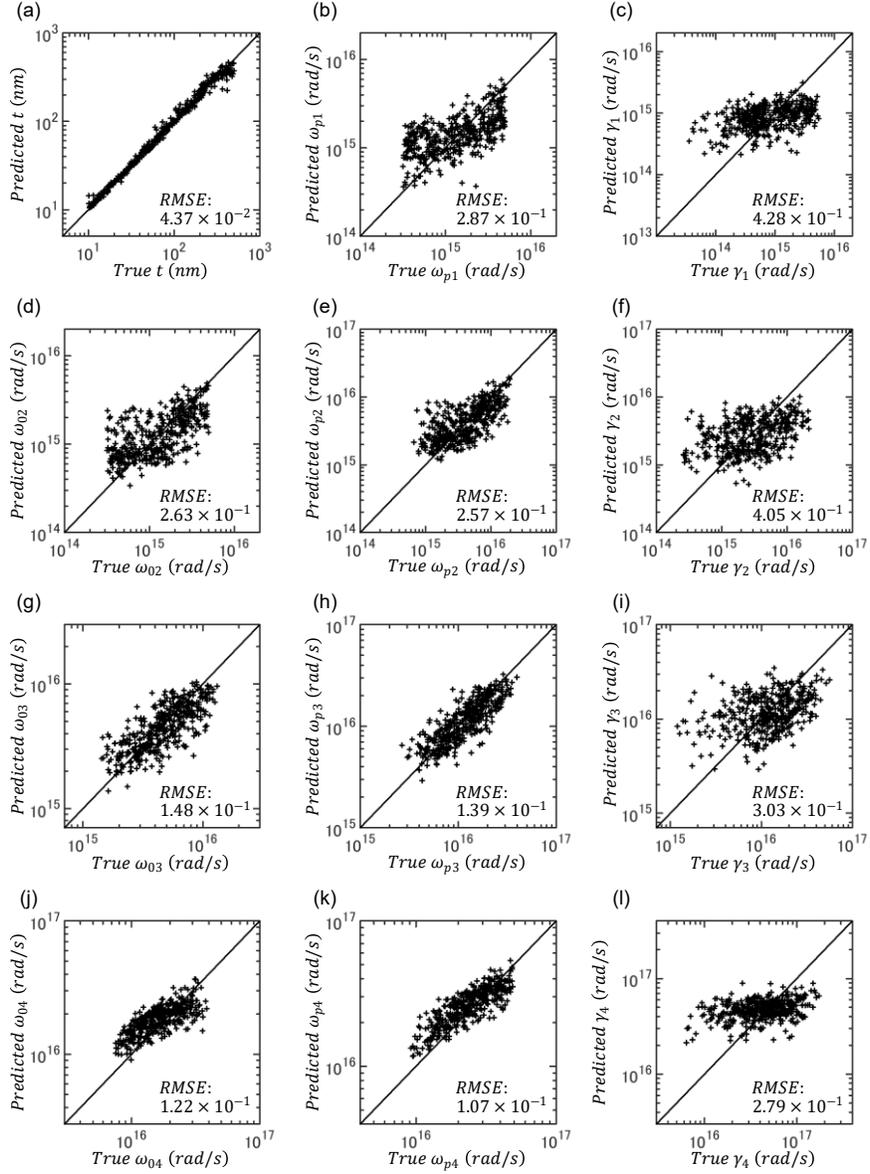

**Figure 4.** Predicted film thickness $t$ and pole parameters in the four-pole model for $\epsilon_d(\omega)$ in Eq. (1). 400 testing incidences are plotted in each panel. RMSE is given at the right bottom of each panel. (a) Film thickness $t$. (b)-(l) Pole parameters in Eq. (1) are shown in (b) for $\omega_{p1}$, (c) for $\gamma_1$, (d) for $\omega_{02}$, (e) for $\omega_{p2}$, (f) for $\gamma_2$, (g) for $\omega_{03}$, (h) for $\omega_{p3}$, (i) for $\gamma_3$, (j) for $\omega_{04}$, (k) for $\omega_{p4}$, and (l) for $\gamma_4$. The variation ranges of those parameters and the hyperparameters for machine learning are presented in Tables S4 and S5 of Supplemental Material [40].

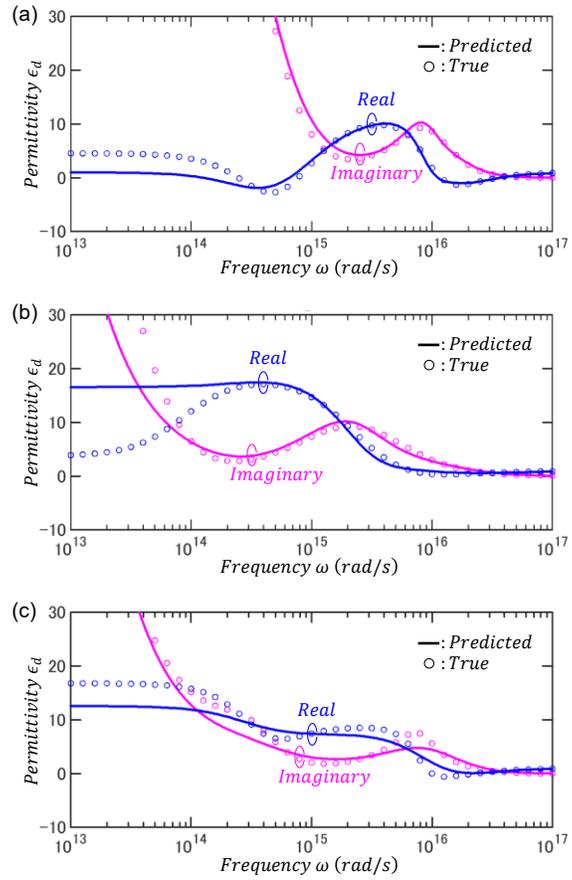

**Figure 5.** (a)-(c) Predicted permittivity spectra for the four-pole model. These examples are selected from the testing data set presented in Figure 4. The blue and pink solid lines represent the real and imaginary parts of predicted $\epsilon_d(\omega)$, and blue and pink symbols represent the real and imaginary parts of true $\epsilon_d(\omega)$. The pole parameters in Eq. (1) for (a)-(c) are presented in Table S6 of Supplemental Material [40].

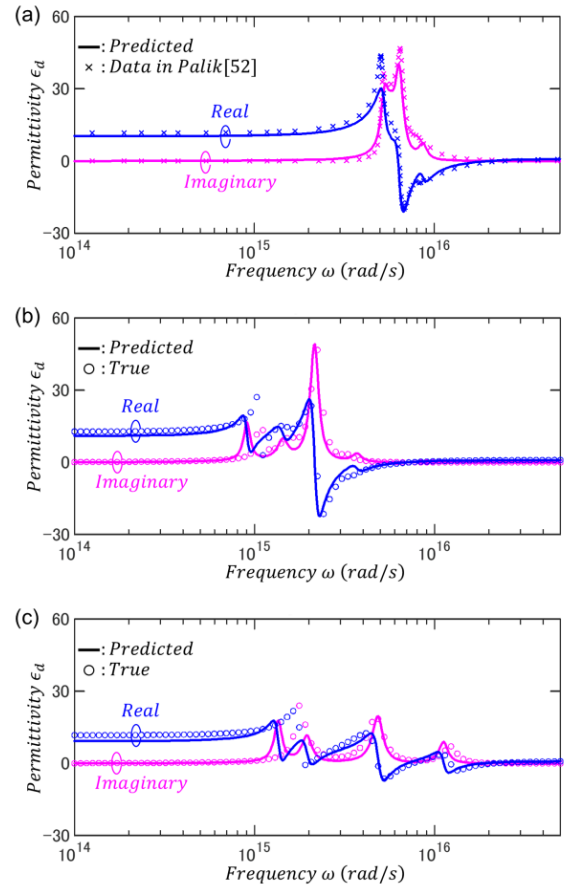

**Figure 6.** (a)-(c) Predicted permittivity spectra with (a) corresponding to silicon. These examples are selected from the testing data set presented in Figure S3 of Supplemental Material [40]. The blue and pink solid lines represent the real and imaginary parts of predicted $\epsilon_d(\omega)$. Blue and pink crosses in (a) and circles in (b) and (c) represent the real and imaginary parts of silicon permittivity in Ref. 52 and true $\epsilon_d(\omega)$, respectively. The pole parameters in Eq. (1) for (a)-(c) are presented in Table S9 of Supplemental Material [40].

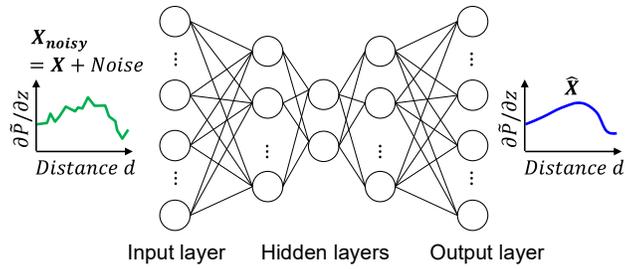

**Figure 7. Denoising autoencoder as the preprocessing for the scheme of Figure 1.** Noisy Casimir forces $X_{noisy}$ with variation of gap distance $d$ are input into the neural network, where the middle hidden layer has a reduced number of neurons. The neural network is trained so that the Casimir forces $\hat{X}$ in the output layer can be close to the clean (target) Casimir forces $X$, i.e. noise effect can be minimized.

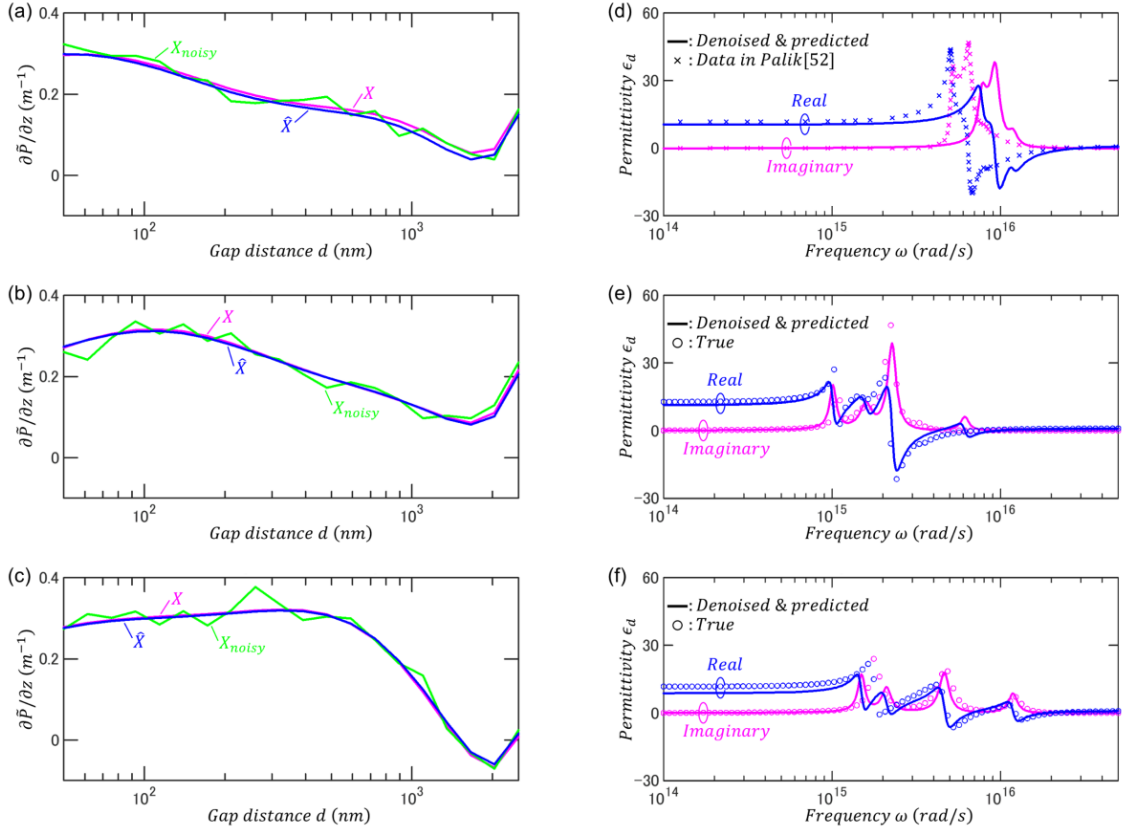

**Figure 8.** (a)-(c) Denoised Casimir forces with variation of gap distance $d$ obtained by using the autoencoder of Figure 7 from Casimir force values corresponding to Figure 6(a)-6(c). Blue, pink, and green lines represent the predicted forces $\widehat{X}$, clean (target) forces $X$, and noisy forces $X_{noisy}$. The hyperparameters for the denoisisng autoencoder are presented in Table S10 of Supplemental Material [40]. (d)-(f) Predicted permittivity spectra from the denoised Casimir forces $\widehat{X}$ of (a)-(c), respectively, using the nerural network of Figure 1. The blue and pink solid lines represent the real and imaginary parts of predicted $\epsilon_d(\omega)$. Blue and pink crosses in (d) and circles in (e) and (f) represent the real and imaginary parts of silicon permittivity in Ref. 52 and true $\epsilon_d(\omega)$, respectively. The pole parameters in Eq. (1) for (d)-(f) are presented in Table S11 of Supplemental Material [40].

# Supplemental material: Machine-learning-assisted material and geometry characterization from Casimir force measurement

Hideo Iizuka[1] and Shanhui Fan[2]
[1]Toyota Central R&D Labs., Inc., Nagakute, Aichi 480 1192, Japan
[2]Department of Electrical Engineering, Ginzton Laboratory, Stanford University,
Stanford, California 94305, United States

**S1. Numerical examples of training data in machine learning**

Numerical examples of training data for the four-pole model are presented. Casimir forces $P$ are calculated from typical three examples (i)-(iii) of film thickness $t$ and pole parameters of the four-pole model for $\epsilon_d(\omega)$ in Table S1, and the normalized Casimir forces $\tilde{P} = \frac{P}{P_P}$ and their derivative to $z$, $\frac{\partial \tilde{P}}{\partial z}$ are shown in Figure S1(a) and S1(b), respectively. The input data in machine learning are expressed by the $L_I$-dimensional vector as

$$X = \left( \left.\frac{\partial \tilde{P}}{\partial z}\right|_{d_1}, \left.\frac{\partial \tilde{P}}{\partial z}\right|_{d_2}, \cdots, \left.\frac{\partial \tilde{P}}{\partial z}\right|_{d_{L_I}} \right)^T, \quad (S1)$$

where those values are taken from Figure S1(b) with an equal step in logscale. Throughout the letter, $L_I = 20$ is selected.

The true values in the output layer are the $L_O$-dimensional vector $Y$, which is expressed as

$$Y = (t, \omega_{01}, \omega_{p1}, \gamma_1, \omega_{02}, \omega_{p2}, \gamma_2, \omega_{03}, \omega_{p3}, \gamma_3, \omega_{04}, \omega_{p4}, \gamma_4)^T, \quad (S2)$$

where $L_O = 13$ for the four-pole model.

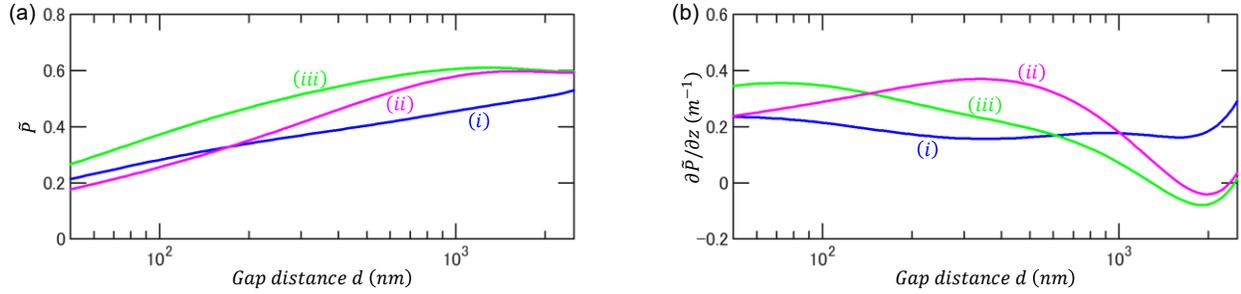

**Figure S1.** Examples of (a) the normalized Casimir forces $\tilde{P} = \frac{P}{P_P}$ and (b) their derivative to $z$, $\frac{\partial \tilde{P}}{\partial z}$ for the four-pole model. Those are calculated from film thickness $t$ and pole parameters of the four-pole model for $\epsilon_d(\omega)$ in Table S1.

**Table S1.** Three sets of film thickness $t$ and pole parameters of the four-pole model for $\epsilon_d(\omega)$.

| Parameter | (i) | (ii) | (iii) |
|---|---|---|---|
| $t (nm)$ | 467 | 167 | 152 |
| $\omega_{01} (10^{15}\ rad/s)$ | 5.51 | 0.746 | 3.19 |
| $\omega_{p1} (10^{15}\ rad/s)$ | 7.67 | 2.08 | 8.03 |
| $\gamma_1 (10^{15}\ rad/s)$ | 0.579 | 0.071 | 0.246 |
| $\omega_{02} (10^{15}\ rad/s)$ | 6.02 | 1.42 | 3.58 |
| $\omega_{p2} (10^{15}\ rad/s)$ | 5.4 | 2.63 | 6.76 |
| $\gamma_2 (10^{15}\ rad/s)$ | 1.5 | 0.288 | 0.734 |
| $\omega_{03} (10^{15}\ rad/s)$ | 15.9 | 13.1 | 4.04 |
| $\omega_{p3} (10^{15}\ rad/s)$ | 20.6 | 13.2 | 15.1 |
| $\gamma_3 (10^{15}\ rad/s)$ | 2 | 2.52 | 0.871 |
| $\omega_{04} (10^{15}\ rad/s)$ | 19.3 | 14.1 | 6.15 |
| $\omega_{p4} (10^{15}\ rad/s)$ | 24.7 | 11.9 | 7.73 |
| $\gamma_4 (10^{15}\ rad/s)$ | 3.78 | 2.33 | 0.418 |

## S2. Hyperparameters for obtaining the prediction results of Figure 2

In the machine learning architecture of Figure 1, all hidden layers have the same number of neurons throughout the paper. The sigmoid function is selected as the activation function. The back-propagation algorithm [S1] is employed. The hyperparameters for obtaining the prediction results of Figure 2 are shown in Table S2.

Table S2. Hyperparameters for obtaining the prediction results of Figure 2.

| Hyperparameter | Value |
|---|---|
| Number of hiddel layers $N_L$ | 3 |
| Number of neurons in each layer $N_N$ | 20 |
| Learning rate | 0.1 |
| Epoch | $5 \times 10^6$ |
| Batch size $M$ | 200 |

## S3. Examples of predicted and true values of film thickness $t$ and resonance frequency $\omega_{02}$ in Figure 2

Table S3 shows two examples of predicted and true values of film thickness $t$ and resonance frequency $\omega_{02}$ in Figure 2. The values of $\omega_{02}$ are used for obtaining permittivity spectra of Figure 3(a) and 3(b). Other parameters of the 1st and 2nd poles are presented in the caption of Figure 2.

Table S3. Examples of predicted and true values of film thickness $t$ and resonance frequency $\omega_{02}$. The values of $\omega_{02}$ are used for obtaining permittivity spectra of Figure 3(a) and 3(b).

| Parameter | Figure 3(a) | | Figure 3(b) | |
|---|---|---|---|---|
| | Predicted | True | Predicted | True |
| $t(nm)$ | 394 | 385 | 63 | 63 |
| $\omega_{02}(10^{15}\ rad/s)$ | 0.366 | 0.365 | 8.65 | 8.74 |

## S4. Variation ranges of parameters for obtaining the prediction results of Figure 4

Table S4 shows the variation ranges of film thickness $t$ and pole parameters for $\epsilon_d(\omega)$ for obtaining the prediction results of Figure 4.

Table S4. Variation ranges of film thickness $t$ and pole parameters for $\epsilon_d(\omega)$ for obtaining the prediction results of Figure 4.

| Parameter | Range |
|---|---|
| $t(nm)$ | $10 \leq t \leq 500$ |
| $\omega_{01}(10^{15}\ rad/s)$ | 0 |
| $\omega_{p1}(10^{15}\ rad/s)$ | $0.316 \leq \omega_{p1} \leq 5$ |
| $\gamma_1(10^{15}\ rad/s)$ | $0.034 \leq \gamma_1 \leq 6.05$ |
| $\omega_{02}(10^{15}\ rad/s)$ | $0.316 \leq \omega_{02} \leq 5$ |
| $\omega_{p2}(10^{15}\ rad/s)$ | $0.657 \leq \omega_{p2} \leq 21.3$ |
| $\gamma_2(10^{15}\ rad/s)$ | $0.131 \leq \gamma_2 \leq 23.3$ |
| $\omega_{03}(10^{15}\ rad/s)$ | $1.33 \leq \omega_{03} \leq 13.9$ |
| $\omega_{p3}(10^{15}\ rad/s)$ | $2.47 \leq \omega_{p3} \leq 39.8$ |
| $\gamma_3(10^{15}\ rad/s)$ | $0.893 \leq \gamma_3 \leq 57.5$ |
| $\omega_{04}(10^{15}\ rad/s)$ | $6.96 \leq \omega_{04} \leq 41.6$ |
| $\omega_{p4}(10^{15}\ rad/s)$ | $7.49 \leq \omega_{p4} \leq 48.9$ |
| $\gamma_4(10^{15}\ rad/s)$ | $5.32 \leq \gamma_4 \leq 198$ |

## S5. Hyperparameters for obtaining the prediction results of Figure 4

The hyperparameters for obtaining the prediction results of Figure 4 are shown in Table S5.

**Table S5.** Hyperparameters for obtaining the prediction results of Figure 4.

| Hyperparameter | Value |
| --- | --- |
| Number of hiddel layers $N_L$ | 3 |
| Number of neurons in each layer $N_N$ | 20 |
| Learning rate | 0.1 |
| Epoch | $4 \times 10^5$ |
| Batch size M | 200 |

## S6. Examples of predicted and true values of film thickness $t$ and pole parameters for $\epsilon_d(\omega)$ in Figure 4

Table S6 shows examples of predicted and true values of film thickness $t$ and pole parameters for $\epsilon_d(\omega)$ in Figure 4. The values of the pole parameters are used for obtaining permittivity spectra of Figure 5(a)-5(c).

**Table S6.** Examples of predicted and true values of film thickness $t$ and pole parameters for $\epsilon_d(\omega)$ in Figure 4. The values of the pole parameters are used for obtaining permittivity spectra of Figure 5(a)-5(c).

| Parameter | Figure 5(a) | | Figure 5(b) | | Figure 5(c) | |
| --- | --- | --- | --- | --- | --- | --- |
| | Predicted | True | Predicted | True | Predicted | True |
| $t(nm)$ | 181 | 194 | 323 | 294 | 47 | 48 |
| $\omega_{01}(10^{15}\ rad/s)$ | 0 | 0 | 0 | 0 | 0 | 0 |
| $\omega_{p1}(10^{15}\ rad/s)$ | 3.76 | 3 | 0.408 | 0.334 | 0.781 | 0.727 |
| $\gamma_1(10^{15}\ rad/s)$ | 0.838 | 0.612 | 0.273 | 0.0871 | 0.561 | 0.463 |
| $\omega_{02}(10^{15}\ rad/s)$ | 0.932 | 0.513 | 2.78 | 4.03 | 0.916 | 0.386 |
| $\omega_{p2}(10^{15}\ rad/s)$ | 2.87 | 2.12 | 8.62 | 13.6 | 2.32 | 1.23 |
| $\gamma_2(10^{15}\ rad/s)$ | 3.05 | 0.978 | 5.16 | 10.5 | 3.08 | 0.635 |
| $\omega_{03}(10^{15}\ rad/s)$ | 9.02 | 9.53 | 6.59 | 7.29 | 10.2 | 8.2 |
| $\omega_{p3}(10^{15}\ rad/s)$ | 25.6 | 26.2 | 16.1 | 12.7 | 21.5 | 19.4 |
| $\gamma_3(10^{15}\ rad/s)$ | 8.91 | 10.8 | 20.7 | 19.8 | 17.4 | 8.02 |
| $\omega_{04}(10^{15}\ rad/s)$ | 17.9 | 14.8 | 22.4 | 22 | 16.7 | 15.4 |
| $\omega_{p4}(10^{15}\ rad/s)$ | 28.7 | 25.2 | 33.4 | 38.5 | 26.9 | 24.2 |
| $\gamma_4(10^{15}\ rad/s)$ | 22.4 | 19.1 | 60.1 | 110 | 30.3 | 31.7 |

## S7. Distance dependencies of Casimir forces calculated from material permittivities and film thicknesses

In the two-pole model with a single parameter $\omega_{02}$, $\omega_{02}$ is widely varied from $1 \times 10^{10} rad/s$ to $1 \times 10^{21} rad/s$, and Casimir forces are calculated from those permittivities and film thicknesses. Figure S2(a) and S2(b) shows the normalized Casimir forces $\tilde{P}$ as a function of vacuum gap distance $d$ for two parameter sets, which correspond to Figure 2(a),(b) and 2(c),(d), respectively. We see that Casimir forces vary strongly with respect to the variation of $\omega_{02}$ in ranges of $1nm \leq d \leq 1000nm$ and $3 \times 10^{14} rad/s \leq \omega_{02} \leq 1 \times 10^{16} rad/s$. In the ranges, we therefore expect that our machine learning approach should work well, as is validated in the results in the main text. On the other hand, the Casimir forces show very little variation when $\omega_{02}$ is increased from $1 \times 10^{16} rad/s$ to $1 \times 10^{21} rad/s$, since there is the cutoff frequency in Casimir forces, above which the Casimir forces are independent of the material permittivity [S2]. Also, the Casimir forces show very little variation when $\omega_{02}$ is decreased from $3 \times 10^{14} rad/s$ to $1 \times 10^{10} rad/s$, which can be understood in the Casimir force formula [Eq. (2)], i.e. the integration is in the form of $\int_0^\infty k_\parallel dk_\parallel$ [S3], as we discuss in the main text. Consequently, if $\omega_{02}$ is outside the range of $3 \times 10^{14} \leq \omega_{02} \leq 1 \times 10^{16} rad/s$, it becomes difficult for our approach to determine $\omega_{02}$ accurately.

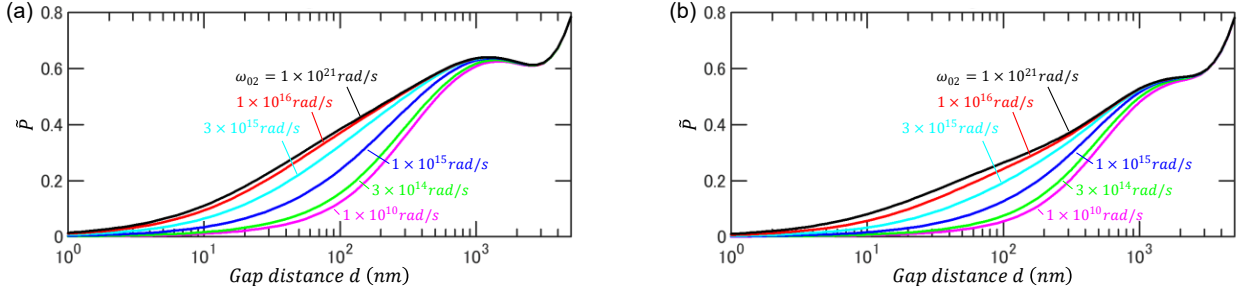

**Figure S2.** Examples of the normalized Casimir forces $\tilde{P} = \frac{P}{P_P}$ as a function of vacuum gap distance $d$ in the two-pole model. $\omega_{02}$ is varied as $1 \times 10^{10} rad/s$ (pink lines), $3 \times 10^{14} rad/s$ (green lines), $1 \times 10^{15} rad/s$ (blue lines), $3 \times 10^{15} rad/s$ (cyan lines), $1 \times 10^{16} rad/s$ (red lines), and $1 \times 10^{21} rad/s$ (black lines). Other parameters are $\omega_{01} = 0$, $\omega_{p1} = 1 \times 10^{15} rad/s$, $\gamma_1 = 1 \times 10^{14} rad/s$, $\omega_{p2} = 4\omega_{02}$, $\gamma_2 = 0.4\omega_{02}$, and $t = 100nm$ in (a), and $\omega_{01} = 0$, $\omega_{p1} = 5 \times 10^{14} rad/s$, $\gamma_1 = 2 \times 10^{14} rad/s$, $\omega_{p2} = 2\omega_{02}$, and $\gamma_2 = 0.8\omega_{02}$, and $t = 200nm$ in (b).

## S8. Variation ranges of parameters for obtaining the prediction results of Figure S3 that is used for permittivity spectra of Figure 6

Table S7 shows the variation ranges of parameters for obtaining the prediction results of Figure S3.

**Table S7.** Variation ranges of parameters for obtaining the prediction results of Figure S3.

| Parameter | Range |
|---|---|
| $t(nm)$ | $100 \leq t \leq 500$ |
| $\omega_{01}(10^{15}\ rad/s)$ | $0.5 \leq \omega_{01} \leq 12.5$ |
| $\omega_{p1}(10^{15}\ rad/s)$ | $0.33 \leq \omega_{p1} \leq 33.4$ |
| $\gamma_1(10^{15}\ rad/s)$ | $0.029 \leq \gamma_1 \leq 2.24$ |
| $\omega_{02}(10^{15}\ rad/s)$ | $0.719 \leq \omega_{02} \leq 13.3$ |
| $\omega_{p2}(10^{15}\ rad/s)$ | $0.462 \leq \omega_{p2} \leq 33.2$ |
| $\gamma_2(10^{15}\ rad/s)$ | $0.04 \leq \gamma_2 \leq 4.07$ |
| $\omega_{03}(10^{15}\ rad/s)$ | $1.01 \leq \omega_{03} \leq 23.6$ |
| $\omega_{p3}(10^{15}\ rad/s)$ | $1.09 \leq \omega_{p3} \leq 46.5$ |
| $\gamma_3(10^{15}\ rad/s)$ | $0.067 \leq \gamma_3 \leq 5.79$ |
| $\omega_{04}(10^{15}\ rad/s)$ | $1.84 \leq \omega_{04} \leq 31.2$ |
| $\omega_{p4}(10^{15}\ rad/s)$ | $1.08 \leq \omega_{p4} \leq 36.7$ |
| $\gamma_4(10^{15}\ rad/s)$ | $0.159 \leq \gamma_4 \leq 6.89$ |

## S9. Hyperparameters for obtaining the prediction results of Figure S3 that is used for permittivity spectra of Figure 6

The hyperparameters for obtaining the prediction results of Figure S3 are shown in Table S8.

**Table S8.** Hyperparameters for obtaining the prediction results of Figure S3.

| Hyperparameter | Value |
|---|---|
| Number of hiddel layers $N_L$ | 3 |
| Number of neurons in each layer $N_N$ | 20 |
| Learning rate | 0.1 |
| Epoch | $8 \times 10^5$ |
| Batch size $M$ | 200 |

## S10. Prediction results for permittivity spectra of Figure 6 (Film permittivities including silicon)

Figure S3 shows the prediction results of the pole parameters of the four-pole model for film permittivity $\epsilon_d(\omega)$ and thickness $t$, using the testing data set generated in the variation ranges of Table S7.

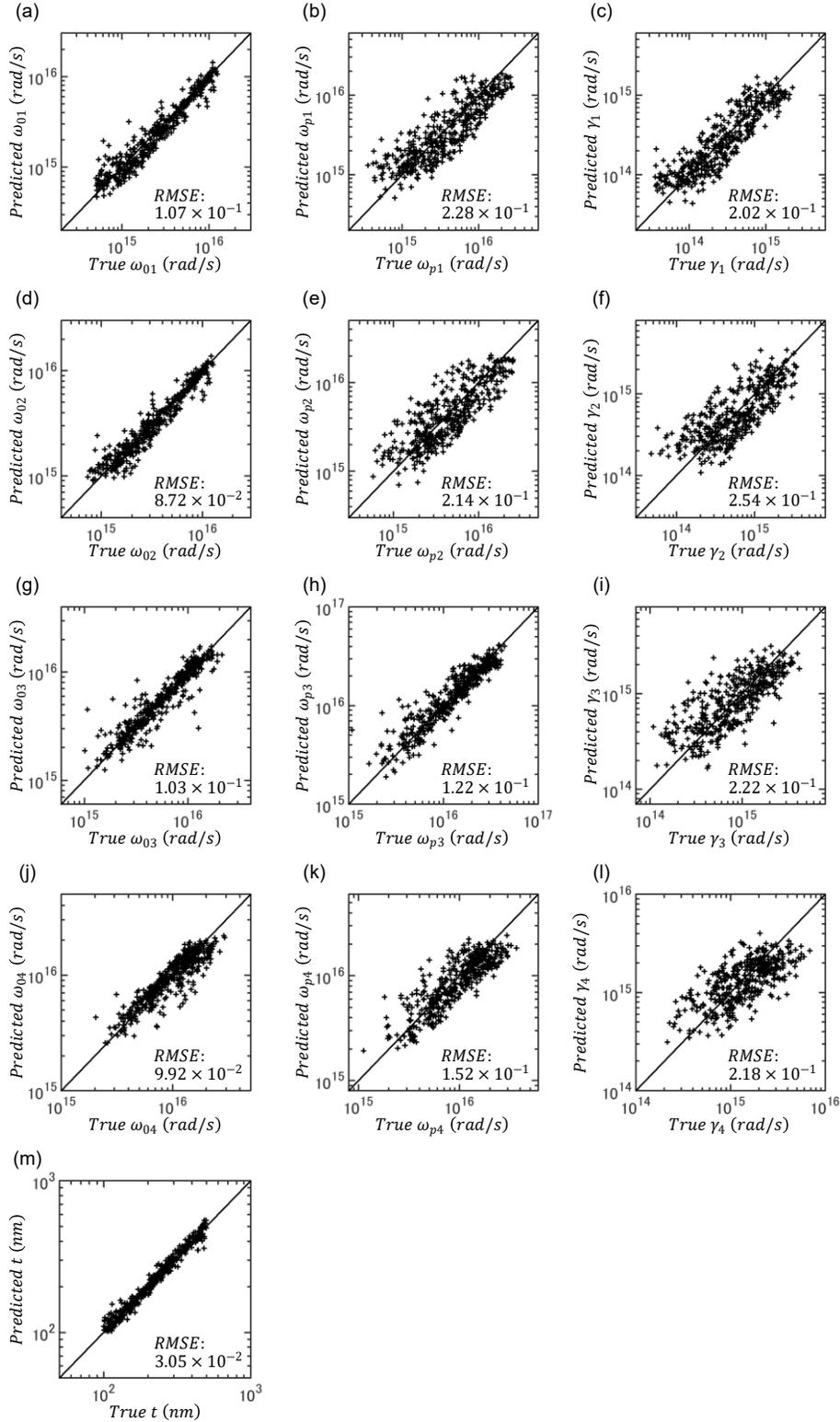

**Figure S3.** Predicted pole parameters for $\epsilon(\omega)$ and film thickness $t$. 400 testing incidences are plotted in each panel. RMSE is given at the right bottom of each panel. (a-l) Pole parameters are shown in (a) for $\omega_{01}$, (b) for $\omega_{p1}$, (c) for $\gamma_1$, (d) for $\omega_{02}$, (e) for $\omega_{p2}$, (f) for $\gamma_2$, (g) for $\omega_{03}$, (h) for $\omega_{p3}$, (i) for $\gamma_3$, (j) for $\omega_{04}$, (k) for $\omega_{p4}$, and (l) for $\gamma_4$. (m) Film thickness $t$. The variation ranges of those parameters and the hyperparameters for machine learning are presented in Tables S7 and S8.

## S11. Examples of predicted and true values of film thickness $t$ and pole parameters for $\epsilon_d(\omega)$ in Figure S3

Table S9 shows examples of predicted and true values of film thickness $t$ and pole parameters for $\epsilon_d(\omega)$ in Figure S3. The values of the pole parameters are used for obtaining permittivity spectra of Figure 6(a)-6(c), where Figure 6(a) corresponds to silicon.

**Table S9.** Examples of predicted and true values of film thickness $t$ and pole parameters for $\epsilon_d(\omega)$ in Figure S3. The values of the pole parameters are used for obtaining permittivity spectra of Figure 6(a)-6(c), where Figure 6(a) corresponds to silicon.

| Parameter | Figure 6(a) (Silicon) | | Figure S6(b) | | Figure S6(c) | |
|---|---|---|---|---|---|---|
| | Predicted | True | Predicted | True | Predicted | True |
| $t(nm)$ | 332 | 324 | 484 | 468 | 140 | 129 |
| $\omega_{01}(10^{15}\ rad/s)$ | 5.3 | 5.2 | 0.909 | 1.07 | 1.36 | 1.8 |
| $\omega_{p1}(10^{15}\ rad/s)$ | 7.81 | 9 | 1.17 | 1.35 | 1.94 | 3.34 |
| $\gamma_1(10^{15}\ rad/s)$ | 0.569 | 0.6 | 0.095 | 0.066 | 0.162 | 0.273 |
| $\omega_{02}(10^{15}\ rad/s)$ | 5.73 | 5.7 | 1.44 | 1.56 | 1.96 | 2.13 |
| $\omega_{p2}(10^{15}\ rad/s)$ | 8.4 | 9 | 1.66 | 2.33 | 2.59 | 2.09 |
| $\gamma_2(10^{15}\ rad/s)$ | 0.906 | 1 | 0.235 | 0.496 | 0.309 | 0.693 |
| $\omega_{03}(10^{15}\ rad/s)$ | 6.39 | 6.45 | 2.16 | 2.22 | 4.85 | 4.76 |
| $\omega_{p3}(10^{15}\ rad/s)$ | 13.2 | 15.4 | 5.46 | 6.1 | 8.71 | 10.5 |
| $\gamma_3(10^{15}\ rad/s)$ | 0.798 | 1 | 0.28 | 0.359 | 0.803 | 1.2 |
| $\omega_{04}(10^{15}\ rad/s)$ | 8.67 | 8.1 | 3.71 | 3.29 | 11.2 | 12.3 |
| $\omega_{p4}(10^{15}\ rad/s)$ | 7.6 | 8 | 2.68 | 1.98 | 12.6 | 15.1 |
| $\gamma_4(10^{15}\ rad/s)$ | 1.09 | 1.2 | 0.565 | 0.564 | 1.58 | 2.62 |

## S12. Hyperparameters for obtaining denoised Casimir forces of Figure 8(a)-8(c) by using the autoencoder

The hyperparameters for obtaining the denoised Casimir forces of Figure 8(a)-8(c) by using the autoencoder of Figure 7 are shown in Table S10.

**Table. S10.** Hyperparameters for obtaining the denoised Casimir forces of Figure 8(a)-8(c) by using the autoencoder.

| Hyperparameter | Value |
|---|---|
| Number of hiddel layers $N_L$ | 3 |
| Number of neurons in each layer | 12,4,12 |
| Learning rate | 0.004 |
| Epoch | $1 \times 10^6$ |
| Batch size M | 200 |

## S13. Predicted and true values of film thickness $t$ and pole parameters for $\epsilon_d(\omega)$ from denoised Casimir forces

Table S11 shows examples of predicted and true values of film thickness $t$ and pole parameters for $\epsilon_d(\omega)$ from denoised Casimir forces of Figure 8(a)-8(c). The corresponding permittivity spectra are shown in Figure 8(d)-S8(f).

**Table. S11.** Predicted film thickness $t$ and pole parameters for $\epsilon_d(\omega)$ from denoised Casimir forces of Figure 8(a)-8(c). The values of the pole parameters are used for obtaining permittivity spectra of Figure 8(d)-8(f). The true values are the same as those of Table S9.

| Parameter | Figure 8(d) (Silicon) | | Figure 8(e) | | Figure 8(f) | |
|---|---|---|---|---|---|---|
| | Denoised & Predicted | True | Denoised & Predicted | True | Denoised & Predicted | True |
| $t(nm)$ | 331 | 324 | 462 | 468 | 148 | 129 |
| $\omega_{01}(10^{15}\ rad/s)$ | 7.77 | 5.2 | 1.01 | 1.07 | 1.5 | 1.8 |
| $\omega_{p1}(10^{15}\ rad/s)$ | 10.4 | 9 | 1.47 | 1.35 | 2.08 | 3.34 |
| $\gamma_1(10^{15}\ rad/s)$ | 0.897 | 0.6 | 0.111 | 0.066 | 0.177 | 0.273 |
| $\omega_{02}(10^{15}\ rad/s)$ | 8.31 | 5.7 | 1.59 | 1.56 | 2.11 | 2.13 |
| $\omega_{p2}(10^{15}\ rad/s)$ | 11.9 | 9 | 2.23 | 2.33 | 2.75 | 2.09 |
| $\gamma_2(10^{15}\ rad/s)$ | 1.71 | 1 | 0.288 | 0.496 | 0.331 | 0.693 |
| $\omega_{03}(10^{15}\ rad/s)$ | 9.32 | 6.45 | 2.28 | 2.22 | 4.66 | 4.76 |
| $\omega_{p3}(10^{15}\ rad/s)$ | 20.6 | 15.4 | 5.29 | 6.1 | 7.92 | 10.5 |
| $\gamma_3(10^{15}\ rad/s)$ | 1.41 | 1 | 0.322 | 0.359 | 0.758 | 1.2 |
| $\omega_{04}(10^{15}\ rad/s)$ | 11.9 | 8.1 | 6.15 | 3.29 | 11.9 | 12.3 |
| $\omega_{p4}(10^{15}\ rad/s)$ | 10.7 | 8 | 5.63 | 1.98 | 13.3 | 15.1 |
| $\gamma_4(10^{15}\ rad/s)$ | 1.71 | 1.2 | 0.841 | 0.564 | 1.7 | 2.62 |